%% file: cdc2018.tex
\documentclass[letterpaper, 10pt, conference]{ieeeconf}    

\IEEEoverridecommandlockouts   
\usepackage[utf8]{inputenc}
\usepackage[T1]{fontenc}
\usepackage[scaled=0.95]{inconsolata}

\usepackage{latexsym,amsmath,amssymb,amsfonts,mathrsfs,mathtools}
\usepackage{arydshln}   

\usepackage{amsthm}
\usepackage{empheq}     

\usepackage{algorithm, algpseudocode}
\makeatletter
\renewcommand{\ALG@beginalgorithmic}{\small}
\makeatother

\usepackage{float, graphicx, caption, subcaption}
\usepackage[usenames,dvipsnames]{xcolor}
\makeatletter
\let\NAT@parse\undefined
\makeatother
\usepackage[colorlinks=true,linkcolor=magenta,citecolor=blue,
            urlcolor=cyan,filecolor=red]{hyperref}
\usepackage{cite}

\newtheorem{theorem}{Theorem}
\newtheorem{lemma}[theorem]{Lemma}
\newtheorem{proposition}[theorem]{Proposition}

\theoremstyle{definition}

\theoremstyle{remark}
\newtheorem{remark}{Remark}

\input{supports/userdef-mathsym.tex}

\hypersetup{
  pdftitle={},
  pdfauthor={Zuogong YUE},
  pdfcreator={Emacs version 25.3.1 + AUCTeX version 11.90}}

\title{\LARGE \bf
  A state-space approach to sparse dynamic network reconstruction
}
\author{Zuogong Yue$^{1*}$, Johan Thunberg$^{1}$, Lennart Ljung$^{2}$ and Jorge Gon\c{c}alves$^{1}$%
  \thanks{This work was supported by Fonds National de la Recherche Luxembourg (Ref.~AFR-9247977, C14/BM/8231540).}%
  \thanks{$^{1}$ Zuogong Yue, Johan Thunberg and Jorge Gon\c{c}alves are with Luxembourg Centre for Systems Biomedicine (LCSB), University of Luxembourg, 7 Avenue des Hauts Fourneaux, 4362, Esch-sur-Alzette, Luxembourg.}%
  \thanks{$^{2}$ Lennart Ljung is with Department of Electrical Engineering, Link\"{o}ping University, Link\"{o}ping, SE-58183, Sweden.}%
  \thanks{\hspace*{0mm}$^{*}$ For correspondence, \href{mailto:zuogong.yue@uni.lu}{\tt zuogong.yue@uni.lu}}
}

\begin{document}

\maketitle
\thispagestyle{empty}
\pagestyle{empty}

\begin{abstract}
  Dynamic network reconstruction has been shown to be challenging due to the
  requirements on sparse network structures and network identifiability.  The
  direct parametric method (e.g., using ARX models) requires a large amount of
  parameters in model selection.  Amongst the parametric models, only a
  restricted class can easily be used to address network sparsity without
  rendering the optimization problem intractable. To overcome these problems,
  this paper presents a state-space-based method, which significantly reduces
  the number of unknown parameters in model selection.  Furthermore, we avoid
  various difficulties arising in gradient computation by using the Expectation
  Minimization (EM) algorithm instead. To enhance network sparsity, the prior
  distribution is constructed by using the Sparse Bayesian Learning (SBL)
  approach in the M-step.  To solve the SBL problem, another EM algorithm is
  embedded, where we impose conditions on network identifiability in each
  iteration. In a sum, this paper provides a solution to reconstruct dynamic
  networks that avoids the difficulties inherent to gradient computation and
  simplifies the model selection.
\end{abstract}


\section{Introduction}
\label{sec:introduction}

Network reconstruction is to infer digraphs or networks that depict interactions
between measured variables. Dynamic network reconstruction refers to a class of
methods in system-theoretic perspective that perform reconstruction by
identifying the underlying network models. It manages to deliver causality
information, and to deal with the transitivity issue (i.e. differing between
$A\!\!\rightarrow\!\!B\!\!\rightarrow\!\!C$ and
$(A\!\!\rightarrow\!\!B\!\!\rightarrow\!\!C, A\!\!\rightarrow\!\!C)$). These
advantages show promising potentials in applications, e.g., detecting critical
genes or regulatory paths that are responsible for diseases from whole genome
data in biomedicine.

There have been many studies on network reconstruction, which may or may not be
entitled in the same way. The most well known topic could be Granger causality
(GC) graphs. The GC graphs are inferred by identifying the vector-autoregressive
(VAR) models (parametric method; based on the ``equivalence'' between GC's
definition and VAR) or by performing statistical tests on conditional
probability independence (non-parametric method; GC's modern definition), e.g.,
\cite{Dahlhaus2003}. These classical methods mainly focus on small-scale
networks. This branch is still active: people keep generalizing this concept,
e.g., \cite{Marinazzo2008,Renault2015}; and try to improve the statistical
tests, e.g., \cite{Dimovska2017}. The work in \cite{Chiuso2012} deserves to be
emphasized, which uses kernel-based system identification methods to identify GC
graphs, and considers sparse network structures, which is particularly useful
for large-scale networks. Bayesian networks is another huge branch that studies
the inference of causal interaction between variables. It defines graphical
models based on conditional probability independence or, equivalently,
\emph{d}-separation, e.g., \cite{Pearl2000,Pearl1988}. Learning methods are
built based on sampling methods or Gaussian approximation in Bayesian
statistics, e.g., \cite{Heckerman2008}. For more methods on network inference
and more comprehensive review, see Chapter~1 in \cite{yue2018}.

This paper adopts a network model for LTI systems, known as \emph{dynamical
  structure function} (DSF)\cite{Goncalves2008}. Loosely speaking, it models
each output variable by a multi-input-single-output (MISO) transfer function
with all the other variables as inputs. Due to its origin in biological
applications, it ``defines'' such a model from state-space models (SSMs), and
its further work studies its realization problems \cite{Yuan2015}. Similar
models are also progressively proposed in \cite{Weerts2015,Dankers2013}, which
were directly presented without starting from state-space models
However, such a derivation from SSMs to DSFs is important, since its feasibility
on realization is the necessity in practice. In terms of graphical
representations, the work in \cite{Chetty2015,Warnick2015} studies different
partial structure representations, points out the differences between
\emph{subsystem structure} and \emph{signal structure}, and discusses the
property of shared hidden states in DSFs.  When considering $D$-matrix in SSMs
(see \eqref{eq:ss-sys}), the well-posedness problem rises as the control theory
\cite{Woodbury2017}. This paper will show how the definition of DSF from SSMs
and its realization contribute to the identification problem. In regard to
network identifiability, the original contribution traces back to
\cite{Goncalves2008} and its successive work, e.g., \cite{Hayden2016a}. The
identification of DSFs shows to be challenging mainly due to the integration of
network identifiability and the imposition of sparse network structure.

\section{Problem Description}
\label{sec:problem-description}

\begin{figure}[htbp]   
  \centering
  \includegraphics[width=.2\textwidth]{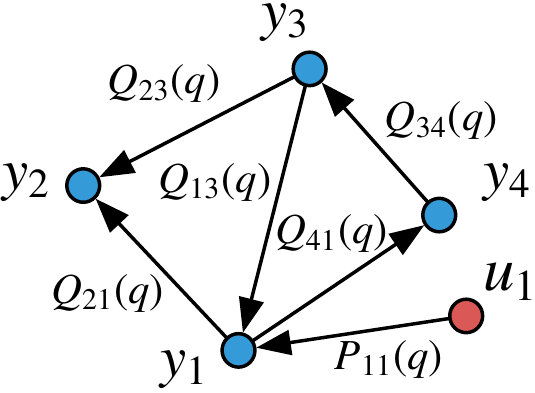}
  \caption{An example of a dynamic network for a given DSF. Here $\mathrm{y}_i$
    denotes the $i$-th element of the output variable $\mathrm{y}$.}
  \label{fig:eg-dynet}
\end{figure}

Consider a dynamical system given by the discrete-time state-space representation in the innovations form
\begin{equation}
  \label{eq:ss-sys}
  \begin{array}{l@{\;}l}
    x(t_{k+1}) &= A x(t_k) + B u(t_k) + K e(t_k),\\
    y(t_{k})   &= C x(t_k) + D u(t_k) +   e(t_k),
  \end{array}
\end{equation}
where $x(t_k)$ and $y(t_k)$ are real-valued $n$ and $p$-dimensional random
variables, respectively; $u(t_k) \in \mathbb{R}^m$, $A, B, C, D, K$ are of
appropriate dimensions; and $\{e(t_k)\}_{k \in \mathbb{N}}$ is a sequence of
i.i.d. $p$-dimensional random variables with $e(t_k)\sim \mathcal{N}(0,R)$. The
initial state $x(t_0)$ is assumed to be a Gaussian random variable with unknown
mean $m_0$ and variance $R_0$.  Without loss of generality, we assume $n \geq p$
and $C$ is of full row rank. For simplicity, we will also use $x_k$ to denote
$x(t_k)$, similarly for $y(t_k), u(t_k), e(t_k)$.
To describe the interconnections between measured variables (i.e. the elements
of $y$), the network model, known as \emph{dynamical structure function} (DSF)
\cite{Goncalves2008}, is derived from \eqref{eq:ss-sys} (see appendix for
details)
\begin{equation}
  \label{eq:dsf-sys}
  y(t_k) = Q(q) y(t_k) + P(q) u(t_k) + H(q) e(t_k),
\end{equation}
where $Q, P, H$ are $p\!\times\!p$, $p\!\times\!m$ and $p\!\times\!p$ matrices
of discrete-time transfer functions, respectively; all diagonal elements of $Q$
are zero; each element of $Q$ (except zeros) is a strictly proper real-rational
transfer function\footnote{Indeed $Q$ can be extended to be proper, instead of
  strictly proper, as the model proposed by Prof.~Paul~van~den~Hof
  (e.g. \cite{Weerts2016,VandenHof2013}). The meaning of causality delivered by
  $Q$ is more clear when $Q$ is strictly proper. In the state-space perspective,
  there are at least two cases presenting proper $Q$'s: 1) when $C$ is not full
  row rank, $Q$ turns to be proper, which, however, can be partitioned into a
  strictly proper block and a block of real number; 2) when introducing
  \emph{intricacy} variables in state-space models \cite{Yeung2011,Yeung2010a},
  which appears as a way to include prior knowledge of the system (such as
  partitions of subsystems), $Q$ could be defined to be proper in general, which
  deserves more studies.}, and elements of $P, H$ are proper.  The \emph{dynamic
  networks} \cite{Yue2017a} that visualize the DSFs are given as
$\mathcal{N} = (\mathcal{G}, f)$, where $\mathcal{G} = (V, E)$ is the
\emph{underlying digraph} that is a 2-tuple of vertex set $V$ and directed edge
set $E$, and $f$ is the \emph{capacity function}, sharing with the same
terminology for \emph{networks} known in the field of graph theory. The $E$ of
$\mathcal{G}$ is determined by checking nonzero elements in $Q, P, H$ and $f$
assigns these elements to the corresponding edges. For example, the dynamic
network of the deterministic DSF (no $H, e$) with
$Q = \bm{0 & 0 & Q_{13} & 0\\ Q_{21} & 0 & Q_{23} & 0 \\ 0 & 0 & 0 & Q_{34} \\
  Q_{42} & 0 & 0 & 0}$ and $P = \bm{P_{11} \\ 0 \\ 0 \\ 0}$ is provided in
Fig.~\ref{fig:eg-dynet}.  The procedure to derive the DSFs from state space
models is called the \emph{definition} of DSFs. And the reverse
procedure--finding a state-space model that can generate the given DSF--is the
\emph{realization} of DSFs (e.g. see \cite{Yuan2015}). The method presented in
this paper relies on both procedures, which shows why it is necessary to study
network models from a state-space perspective.

\noindent\textbf{Problem}:
Consider the special case with $R = I$, $K = \sigma \in \mathbb{R}_+$, and
$C = \bm{I & 0}$, $D = 0$.  Let $Y^N \triangleq \{y(t_1), \dots, y(t_N)\}$
denote the measured samples and $U^N \triangleq \{u(t_1), \dots, u(t_N)\}$ be
the input signals, which is assumed to known or measured without input
measurement noise. The dynamic network reconstruction problem is to infer
$\mathcal{N}$ from $(Y^N, U^N)$ by identifying the DSFs \eqref{eq:dsf-sys},
assuming the ground truth network is sparse.

\begin{figure}[htbp]   
  \centering
  \includegraphics[width=.45\textwidth]{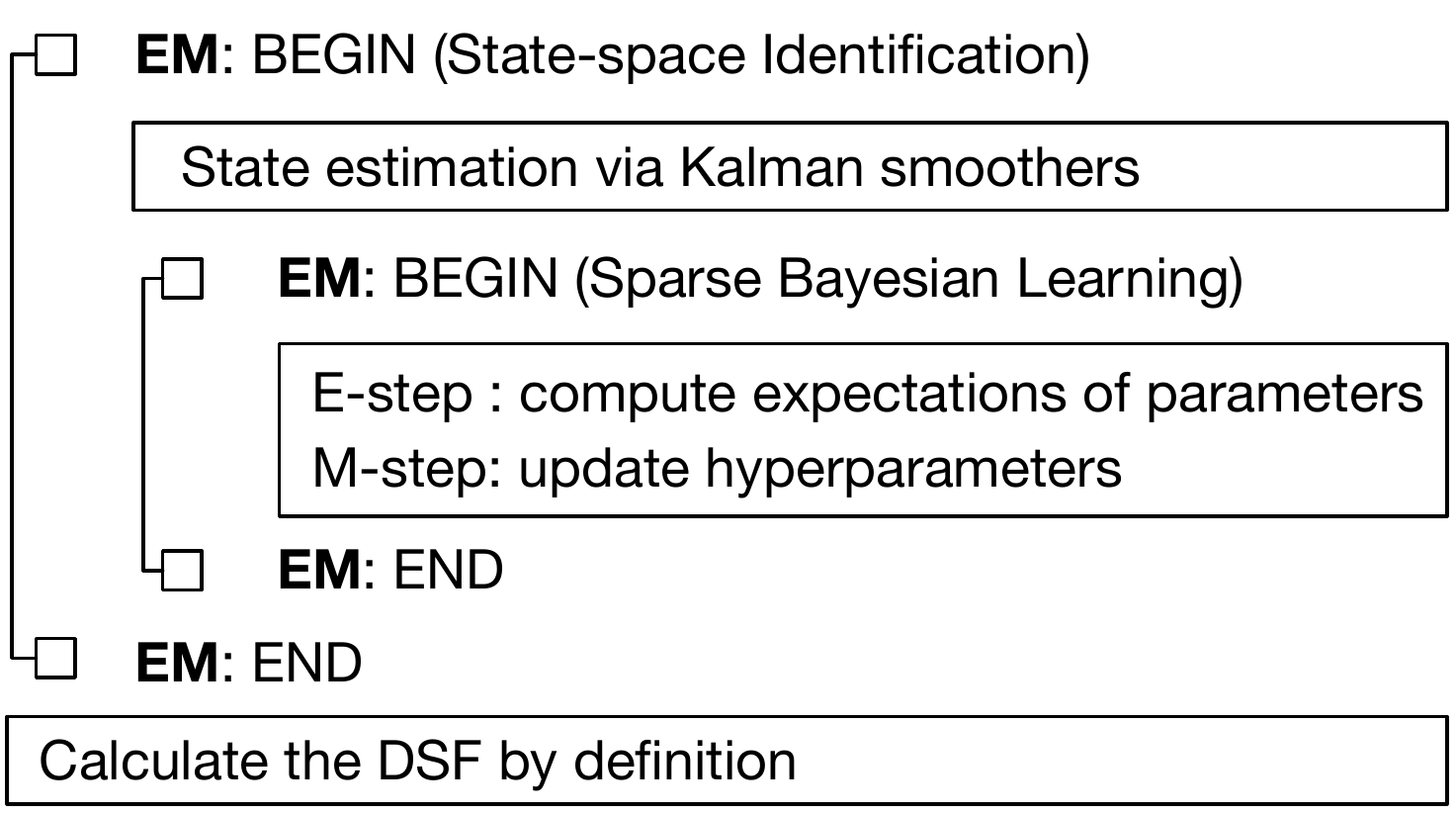}
  \caption{An overview of the state-space-based reconstruction method.}
  \label{fig:overview-algor}
\end{figure}

The method presented in this paper estimate the DSF by identifying a state-space
realization of \eqref{eq:dsf-sys} with the network identifiability guaranteed.
During the reconstruction procedure, we have to take into account the issue of
network identifiability, i.e. whether it is possible to uniquely determine
$(Q,P,H)$ from the input-output data, which guarantees that any state-space
realization leads to the same DSF. The proposed algorithm is illustrated by
Fig.~\ref{fig:overview-algor}, which consists of two EM loops: the outer loop is
the regular EM method for state-space identification; the inner loop performs
the \emph{sparse Bayesian learning} (SBL) for network sparsity and integrates
network identifiability conditions.
In the following two sections we will describe the method in
details. Section~\ref{sec:expect-maxim-algor} describes the outer EM loop,
Section~\ref{sec:sparse-bayes-estim} describes the inner EM loop, and
Section~\ref{sec:integ-netw-ident} describes the integration of network
identifiability in the inner loop.

\begin{remark}
  We restrict the state-space models to be the innovations forms, considering
  the availability of a definition of dynamical structure functions.  It is well
  known that a general LTI state-space model (with different process and
  measurement noises) can be transformed into the innovations form. However, we
  have not presented a definition of DSFs that is invariant to such a
  transformation. Moreover, we only consider the particular case
  $Ke(t_k) \sim \mathcal{N}(0, \sigma^2 I)$ due to the restrictions in sparse
  Bayesian learning. Nevertheless, $e(t_k)$ is not necessarily a standard normal
  distribution, whose covariance can be scaled by a positive value.  The
  restrictions of $C=\bm{I & 0}$ and $D = 0$ allow us to directly use results on
  network identifiability in \cite{Hayden2016a}. As shown in the appendix, the
  state transformations due to general $C$'s or nonzero $D$'s in
  \eqref{eq:qph-general-dsf} would invalidate the results presented in
  \cite{Hayden2016a}. However, the extension of \cite{Hayden2016a} is possible
  but deserves more considerations.
\end{remark}

\section{Expectation Maximization}
\label{sec:expect-maxim-algor}

In the ``outer-loop'' EM algorithm, we choose $z(t)$ as the latent variable,
whose values at $t_1, \dots, t_N$ are the ``missing data'', denoted by
$Z^N \triangleq \{z_1, \dots, z_N\} = \bm{0 & I_{n \times p}}X^N$, where
$X^N \triangleq \{x_1, \dots, x_N\}$ is the ``complete data''. Define the
\emph{complete data log likelihood} to be
\begin{math}
  \log p(X^N | \theta)
    = \log p(x_1|\theta) + \sum_{k=2}^N \log p(x_k | x_{k-1}, \theta),
\end{math}
where $p(X^N|\theta)$ is the joint p.d.f. of $x_1, \dots, x_N$ given
$\theta$; $p(x_k | x_{k-1}, \theta)$ is the conditional p.d.f. of $x_k$
given $\theta$ and the previous state value $x_{k-1}$.  However, this cannot
be computed directly due to the lack of the measurements of $z(t)$. Hence, the
E-step is to calculate the \emph{expected complete data log likelihood} using
the available observations $Y^N$
\begin{equation}
  \label{eq:def-expect-complete-data-log-likelihood}
  \begin{array}{l@{\;}l}
    \mathcal{Q}(\theta, \theta^i) &= \mathbb{E}_{\theta^k}\left( \log p(X^N|\theta) \,| Y^N \right)\\
     &\displaystyle\coloneqq \int \log p(Y^N, Z^N | \theta)\: p(Z^N|Y^N, \theta^i) \mathrm{d} Z^N,
  \end{array}
\end{equation}
where $\mathbb{E}_{\theta^i}(\cdot | Y^N)$ denotes the conditional expectation
given the measurements $Y^N$ and the parameters $\theta^i$ (to be more clear,
the alternative notation is $\mathbb{E}_{Z^N|Y^N; \theta^i}(\cdot)$), and $i$
denotes the iteration index.  Instead of maximizing likelihood, we choose the
maximum a posteriori (MAP), which includes an prior distribution that imposes
sparsity of certain parameters and leads to sparse network structure. Let the
whole parameter vector $\theta$ be categorized into two groups: one consists of
unknown deterministic variables including $m_0, R_0, \sigma$; the other
comprises $A, B$, denoted by $\mathrm{w}$, which is treated as a random
variable that commits a prior distribution.  In the M-step, we perform the MAP
to update $\theta$:
\begin{equation}
  \label{eq:MAP-estimate-in-M-step}
  \theta^{i+1} = \argmax{\theta} \mathcal{Q}(\theta, \theta^{i}) + \log
  p(\mathrm{w}, {\gamma}),
\end{equation}
where $p(\mathrm{w}, {\gamma})$ is the prior that depends on hyperparameter
${\gamma}$. The specific construction of $p(\mathrm{w}, {\gamma})$ is presented
in Section~\ref{sec:sparse-bayes-estim}.  Note that we will not maximize
\eqref{eq:MAP-estimate-in-M-step} directly, which is presented merely to show
the principle.

\begin{remark}
  Note that the prior in \eqref{eq:MAP-estimate-in-M-step} might vary over
  iterates due to the update of hyperparameters, which are determined by
  maximizing the \emph{marginal likelihood function} (a.k.a. \emph{evidence
    function}).  It could be problematic to directly cite the usual result that
  the EM algorithm monotonically increases the log posterior of the observed
  data until it reaches a local optimum (see \cite[Sec.~11.4]{Murphy2012}),
  which uses a fixed prior distribution. The convergence deserves more
  discussions in further theoretical studies.
\end{remark}

Now we show the calculation of $\mathcal{Q}(\theta, \theta')$ in
Lemma~\ref{lemma:expect-compl-data-log-likelihood}.  Let
$\hat{x}_{k|l} \triangleq \hat{x}(t_k|t_l)$ denote the estimate of $x(t_k)$
given $x(t_l)$, similarly for $y, z, u$.

\begin{lemma}
  \label{lemma:expect-compl-data-log-likelihood}
  The expected complete data log likelihood $\mathcal{Q}(\theta, \theta')$ is given as follows
  (neglecting the constant terms)
  \begin{equation}
    \label{eq:expect-compl-data-log-likelihood}
    \begin{array}{l@{\;}l}
      &-2 \mathcal{Q}(\theta, \theta')
       = \log\det R_0 + N \log\det (\sigma^2I) \\
      &\hspace*{2mm}+ \Tr \Big\{ R_0^{-1} \mathbb{E}_{\theta'} \big( (x_0 - m_0)(x_0 - m_0)^T | Y^N
      \big) \Big\} \\[1mm]
      &\hspace*{2mm}+\,\sum_{k=1}^N \Tr \Big\{ \sigma^{-2} \big[
        \mathbb{E}_{\theta'}(x_k x_k^T | Y^N) -
        L \mathbb{E}_{\theta'}(x_k \xi_k^T | Y^N)^T\\[1mm]
      &\hspace*{15mm}
      -\,\mathbb{E}_{\theta'}(x_k\xi_k^T | Y^N)L^T +
        L \mathbb{E}_{\theta'}(\xi_k\xi_k^T | Y^N) L^T
      \big] \Big\},
    \end{array}
  \end{equation}
  where $\xi_k \triangleq \bm{x_{k-1}^T & u_{k-1}^T}^T$,
  $L \triangleq \bm{A & B}$.
\end{lemma}
The proof follows straightforwardly by referring to \cite{Gibson2005} and is
hence omitted. The expectations in
Lemma~\ref{lemma:expect-compl-data-log-likelihood} can be computed via Kalman
smoothers, listed in
Lemma~\ref{lemma:expectation-items-for-compl-data-log-likeli}.  The expression
in Lemmar~\ref{lemma:expect-compl-data-log-likelihood} implies that the
maximization of $\mathcal{Q}(\theta,\theta')$ can be split into two parts:
maximizing the part in terms of $A, B, \sigma$; and update the estimations of
$x_0$ and $R_0$ by
$\hat{x}_0 = \mathbb{E}_{\theta'}(x_0 | Y^N), \hat{R}_0 =
\mathbb{E}_{\theta'}(x_0 x_0^T|Y^N)$. One may refer to Lemma~3.3 in
\cite{Gibson2005} for details. Let us denote
$\mathcal{Q}(\theta,\theta') = \mathcal{Q}_1(x_0, R_0) +
\mathcal{Q}_2(\mathrm{w}, \sigma^2)$ (neglecting $\theta'$, which is given in
each iteration).  In favor of the sparsity of $\mathrm{w}$ (i.e. $A, B$), we
will maximize the expected posterior distribution
$Q_2(\mathrm{w, \sigma^2}) + \log p(\mathrm{w},\gamma)$ via sparse Bayesian
learning; and estimate $x_0, R_0$ in the same way.

To close this section, we present the results of the following items:
$\mathbb{E}_{\theta'}(x_k x_k^T|Y^N)$,
$\mathbb{E}_{\theta'}(x_k x_{k-1}^T|Y^N)$,
$\mathbb{E}_{\theta'}(x_k u_{k-1}^T|Y^N)$ and
$\mathbb{E}_{\theta'}(x_{k-1} u_{k-1}^T|Y^N)$ calculated by Kalman
smoothing. Note that
$\mathbb{E}_{\theta'}(u_{k-1} u_{k-1}^T|Y^N) = u_{k-1} u_{k-1}^T$ since $u(t)$
is assumed to be deterministic\footnote{In experiments we can be stochastic
  signals (e.g., the popular Gaussian i.i.d.)  as $u(t)$ to stimulate
  systems. However, since $u(t)$ is assumed to be accessible in identification,
  it is still treated as deterministic.}.

\begin{lemma}
  \label{lemma:expectation-items-for-compl-data-log-likeli}
  Let the parameter vector $\theta'$ be composed of the elements of
  $A, B, C, D, \sigma, m_0, R_0$, which are defined in \eqref{eq:ss-sys}. Then
  \begin{align}
    \label{eq:Expectation-xk-xk}
    \mathbb{E}_{\theta'}(x_k x_k^T|Y^N) &= \hat{x}_{k|N}\hat{x}_{k|N}^T + P_{k|N},\\
    \label{eq:Expectation-xk-xk-1}
    \mathbb{E}_{\theta'}(x_k x_{k-1}^T|Y^N) &= \hat{x}_{k|N}\hat{x}_{k-1|N}^T + M_{k|N},\\
    \label{eq:Expectation-xk-uk-1}
    \mathbb{E}_{\theta'}(x_k u_{k-1}^T|Y^N) &= \hat{x}_{k|N} u_{k-1}^T,\\
    \label{eq:Expectation-xk-1-uk-1}
    \mathbb{E}_{\theta'}(x_{k-1} u_{k-1}^T|Y^N) &= \hat{x}_{k-1|N} u_{k-1}^T,
  \end{align}
  where $\hat{x}_{k|N}, P_{k|N}$ and $M_{k|N}$ are calculated in reverse-time recursions
  via the Kalman smoother
  \begin{equation}
    \label{eq:kalman-smoothering}
    \begin{aligned}
      J_k     &= P_{k|k} A^T P_{k+1|k}^{-1}, \\
      \hat{x}_{k|N} &= \hat{x}_{k|k} + J_k ( \hat{x}_{k+1|N} - \hat{x}_{k+1|k} ), \\
      P_{k|N} &= P_{k|k} + J_k (P_{k+1|N} - P_{k+1|k}) J_k^T,
    \end{aligned}
  \end{equation}
  for $k = N, \dots, 1$, and the lag-one covariance smoother
  \begin{equation}
    \label{eq:lag-one-covariance-smoother}
    M_{k|N} = P_{k|k} J_{k-1}^T + J_{k} (M_{k+1|N} - A P_{k|k}) J_{k-1}^T
  \end{equation}
  for $k = N,\dots,2$.  The quantities $\hat{x}_{k|k}, P_{k|k}, P_{k|k-1}$ and initial
  conditions $\hat{x}_{N|N}, P_{N|N}$ required in
  \eqref{eq:kalman-smoothering},~\eqref{eq:lag-one-covariance-smoother} are
  computed by the Kalman filter
  \begin{equation}
    \label{eq:kalman-filter}
    \begin{aligned}
      \hat{x}_{k|k-1} &= A \hat{x}_{k-1|k-1} + B u_{k-1},\\
      P_{k|k-1} &= A P_{k-1|k-1} A^T + \sigma^2 I, \\
      \hat{x}_{k|k} &= \hat{x}_{k|k-1} + K_k (y_k - C \hat{x}_{k|k-1} - D u_{k}), \\
      P_{k|k} &= P_{k|k-1} - K_k C P_{k|k-1}, \\
      K_k &= P_{k|k-1} C^T (C P_{k|k-1} C^T + I)^{-1},
    \end{aligned}
  \end{equation}
  for $k = 1,\dots,N$. The lag-one covariance smoother is initialized with
  \begin{equation}
    \label{eq:lag-one-cov-smoother-initialization}
    M_{N|N} = (I - K_N C) A P_{N-1|N-1}.
  \end{equation}
\end{lemma}

\section{Sparse Bayesian Learning}
\label{sec:sparse-bayes-estim}

This section describe how we use sparse Bayesian learning in the inner loop
of the proposed method to get sparse solutions. In regard to network sparsity,
we focus on $Q$ in \eqref{eq:dsf-sys}\footnote{The sparsity of $P, H$ in the
  DSFs probably might also be interesting in particular applications. However,
  due to network identifiability, up to the conditions available, we have to
  require either $P$ or $H$ to be diagonal.}. However, here we fail to directly
apply sparsity constraints on $Q$. Instead, we achieve it heuristically by
imposing sparsity requirements on $A$. However, it deserves to be pointed out
that it is possible that $A$ is not sparse but $Q$ is sparse, which case fails
to be covered in this work.

In the setup of the SBL, we treat $x_0$ as a given value. The complete-data
likelihood function is
\begin{equation}
  \label{eq:SBL-likelihood}
  \begin{array}{l@{\:}l}
    p(X^N|\theta) &\triangleq p(x_N, x_{N-1},\dots,x_1|\theta) \\
                    &=p(x_N|x_{N-1},\theta)\cdots p(x_2|x_1,\theta) p(x_1,\theta),
  \end{array}
\end{equation}
where
$p(x_k|x_{k-1},\theta) = \mathcal{N}(x_k - A x_{k-1} - B u_{k-1},
\sigma^2I)$, $k=1,\dots, N$.  We rewrite the complete-data likelihood
$p(X^N|\theta)$ as follows,
\begin{equation}
  \label{eq:SBL-likelihood}
  \begin{array}{l@{\:}l}
    p(\mathrm{y}| \mathrm{w}; \sigma) =
    &(2\pi)^{-N_\mathrm{y}/2} \, |\Sigma|^{-1/2}\\
    & \times \exp \left[ -\frac{1}{2} (\mathrm{y} - \Phi \mathrm{w})^T \Sigma^{-1} (\mathrm{y} - \Phi \mathrm{w}) \right],
  \end{array}
\end{equation}
where
\begin{equation*}
  \mathrm{y} =
  \begin{bmatrix}
    x_N \\ \vdots \\ x_1
  \end{bmatrix},
  \quad
  \Phi =
  \begin{bmatrix}
    \Phi_N \\ \vdots \\ \Phi_1
  \end{bmatrix},
  \;
  \Phi_k =
  \begin{bmatrix}
    x_{k-1}^T\otimes I &
    u_{k-1}^T\otimes I
  \end{bmatrix},
\end{equation*}
\begin{equation*}
  \mathrm{w} =
  \begin{bmatrix}
    \kvec(A) \\ \kvec(B)
  \end{bmatrix},
  \quad
  \Sigma = \blkdiag(\sigma^2I,\dots,\sigma^2I) \; \text{($N$ blocks)},
\end{equation*}
$N_\mathrm{y}$ denotes the dimension of $\mathrm{y}$, and here $I$ denotes the
$n \times n$ identity matrix.  The parameter vector $\theta$ is composed of
$\mathrm{w}$ and $\Sigma$, where $\mathrm{w}$ assumes a parametrized prior and
$\Sigma$ is treated as a deterministic parameter.

As studied in SBL, we introduce the Gaussian prior to impose sparsity on
$\mathrm{w}$,
\begin{equation}
  \label{eq:SBL-prior}
  \begin{array}{l@{\:}l}
    p(\mathrm{w}; {\gamma}) =
    &(2\pi)^{-N_{\mathrm{w}}/2}\, | \Gamma |^{-1/2}\\
    &\times \exp \left(  -\frac{1}{2} \mathrm{w}^T \Gamma^{-1} \mathrm{w}
  \right),
  \end{array}
\end{equation}
where $\Gamma = \diag({\gamma})$\footnote{For convenience, we will use
  $\Gamma$ and ${\gamma}$ interchangeably when appropriate.}, and
$N_\mathrm{w}$ denotes the dimension of $\mathrm{w}$.  The posterior and
marginal likelihood can be obtained in the procedure given by
\cite{Tipping2001}.  For fixed values of the hyperparameters, the complete-data
posterior is Gaussian,
\begin{equation}
  \label{eq:SBL-posterior}
  \begin{array}{l@{\:}l}
    p(\mathrm{w}|\mathrm{y}; {\gamma},\sigma) =
    &(2\pi)^{-N_\mathrm{w}/2}\, |\Sigma_{\mathrm{w}}|^{-1/2} \\
    &\times \exp \left[ -\frac{1}{2} (\mathrm{w} - {\mu}_\mathrm{w})^T \Sigma_\mathrm{w}^{-1}
    (\mathrm{w} - {\mu}_\mathrm{w}) \right],
  \end{array}
\end{equation}
where
\begin{equation}
  \label{eq:SBL-posterior--mean-cov}
  \begin{array}{l@{\:}l}
    {\mu}_\mathrm{w} &= \Sigma_\mathrm{w} \Phi^T \Sigma^{-1} \mathrm{y}, \\
    \Sigma_\mathrm{w} &= \left( \Gamma^{-1} + \Phi^T \Sigma^{-1} \Phi \right)^{-1}.
  \end{array}
\end{equation}
And the marginal likelihood is given by
\begin{equation}
  \label{eq:SBL-marginal}
  p(\mathrm{y}; {\gamma},\sigma) = (2\pi)^{-N_\mathrm{y}/2} |\Sigma_\mathrm{y}|^{-1/2}
  \exp \left( \textstyle-\frac{1}{2} \mathrm{y}^T \Sigma_\mathrm{y}^{-1} \mathrm{y} \right),
\end{equation}
where $\Sigma_\mathrm{y} = \Sigma + \Phi \Gamma \Phi^T$.  The hyperparameters
${\gamma}$ and $\sigma^2$ will be determined by Evidence Maximization or
Type-II Maximum Likelihood \cite{Tipping2001}.

Now the onus remains in estimating ${\gamma}$ and $\sigma$ via evidence
maximization of the complete-data marginal likelihood
\eqref{eq:SBL-marginal}. We employ another EM algorithm to maximize
${p}(\mathrm{y}; {\gamma},\sigma)$, which is equivalent to minimizing
$-2\log{p}(\mathrm{y}; {\gamma},\sigma)$. This EM algorithm proceeds by
choosing $\mathrm{w}$ as the latent variable and minimizing
\begin{equation}
  \label{eq:SBL-EM-compl-data-likelihood}
  \mathbb{E}_{\mathrm{w}|\mathrm{y};{\gamma},\sigma} \left(
    -2\log{p}(\mathrm{y}, \mathrm{w}; {\gamma},\sigma)
  \right),
\end{equation}
where
${p}(\mathrm{y}, \mathrm{w}; {\gamma},\sigma) = {p}(\mathrm{y} | \mathrm{w};
\sigma) {p}(\mathrm{w}; {\gamma})$.  Instead of calculating
\eqref{eq:SBL-EM-compl-data-likelihood} throughout, we calculate certain
expectations in the E-step according to the demands of updating ${\gamma}$ and
$\sigma$ in the M-step. The evidence maximization is performed by computing the
following in the $k$-th iteration:\\[1ex]
\hspace*{1ex} \underline{E-step}:
\begin{equation}
  \label{eq:SBL-alg-E-step}
  \begin{array}{ll}
    &\mathbb{E}_{\mathrm{w}|\mathrm{y};{\gamma}^k,\sigma^k}(\mathrm{w})
      = \big({{\mu}}_\mathrm{w} \big)
      \big|_{{\gamma} = {\gamma}^k, \sigma = \sigma^k},\\[1.5mm]
    &\mathbb{E}_{\mathrm{w}|\mathrm{y};{\gamma}^k,\sigma^k}(\mathrm{w}\mathrm{w}^T)
      = \big( {\Sigma}_\mathrm{w} + {{\mu}}_\mathrm{w}
      {{\mu}}_\mathrm{w}^T \big)
      \big|_{{\gamma} = {\gamma}^k, \sigma = \sigma^k},
  \end{array}
\end{equation}
\hspace*{1ex} \underline{M-step}:
\begin{equation}
  \label{eq:SBL-alg-M-step}
  \begin{array}{ll@{}l@{\:}l}
    & \Gamma^{j+1} &=& \argmini{{\gamma}} \mathbb{E}_{\mathrm{w}| \mathrm{y};
                     {\gamma}^j, \sigma^j} \left(
                     -2\log\hat{p}(\mathrm{y}, \mathrm{w}; {\gamma},\sigma^j)
                     \right) \\
    & &=& \mathbb{E}_{\mathrm{w} | \mathrm{y};{\gamma}^j,\sigma^j}
        (\mathrm{w}\mathrm{w}^T) ,\\[1.5mm]
    & (\sigma^2)^{j+1} &=& \argmini{{\gamma}} \mathbb{E}_{\mathrm{w}| \mathrm{y};
                     {\gamma}^j, \sigma^j} \left(
                     -2\log\hat{p}(\mathrm{y}, \mathrm{w}; {\gamma}^{j},\sigma)
                           \right) \\
    & &=& \frac{1}{N} \left( \|\mathrm{y} - \Phi {\mu}_{\mathrm{w}}\|_2^2 +
          (\sigma^2)^{j} \Tr(I - \Sigma_{\mathrm{w}} \Gamma^{-j})\right),
  \end{array}
\end{equation}
where $j$ denotes the iteration index, $(\sigma^2)^j$  the square of $\sigma^j$,
$\Gamma^{-j}$ the inverse of $\Gamma^j$, and $\Tr(X)$ the trace of matrix $X$.

It is theoretically possible to compute the expectations of
$p(\mathrm{y}|\mathrm{w};\sigma)$ and $p(\mathrm{w};{\gamma})$ via Kalman
smoothers. However, we failed to get closed forms to calculate the expectations
of $p(\mathrm{w}|\mathrm{y};{\gamma}, \sigma)$ and
$p(\mathrm{y};{\gamma},\sigma)$ via Kalman filters and smoothers. Instead, We
approximately compute ${\mu}_\mathrm{w}, \Sigma_\mathrm{w}$ using the state
estimation $\hat{x}_{k|N}$ from Kalman smoothers. The optimal variance
estimation is hence omitted.

Without considerations on the computational cost, there might be better
alternatives for computing ${\mu}_\mathrm{w}$ and $\Sigma_\mathrm{w}$ using
sampling methods. One way is to use particle filters to sample from the
distribution of the complete data $x(t)$, and then use the samples to estimate
${\mu}_\mathrm{w}, \Sigma_\mathrm{w}$. Alternatively, in the special case of
Gaussian $x(t)$ (e.g. no inputs $u(t)$), we could keep using Kalman filters and
smoothers, which provide the optimal estimation of the mean and covariance of
$x(t)$. Then we directly sample from the optimally estimated distribution of
$x(t)$ and compute ${\mu}_\mathrm{w}, \Sigma_\mathrm{w}$ in the Monte Carlo
way.

\section{Integration of Network Identifiability}
\label{sec:integ-netw-ident}

There are two practical ways to guarantee the identifiability. One is to perturb
each output variable by designed signals \cite{Goncalves2008}, and, as a result,
$Q(s)$ in the DSF is inferred using the signals responding to the input
signals. The other is to taking advantages of noises. We need to accept \emph{a
  priori} that the noise model is \emph{minimum-phase}, one underlying
realization of the DSF is \emph{global minimal} and the i.i.d. process noises
perturb the outputs in such a way that $H(s)$ is square, diagonal and full-rank
(see \cite{Hayden2016a}). $Q(s)$ in the DSF is actually inferred from the
responses of process noises. The benefit of the latter is the decrease of the
number of experiments (i.e. $P(s)$ is no longer required to be square). However,
in practice, it is hard to test if we could accept these \emph{a priori}
assumptions on process noises. What's more, the conditions guaranteeing diagonal
$H$ turn to be particular complicated when including measurement noise, e.g. as
shown by \eqref{eq:qph-general-dsf}.  Hence, in this paper, we focus on the
first case--guaranteeing network identifiability via external inputs.

Referring to Theorem~2 in \cite{Goncalves2008}, if $P(s)$ is square, diagonal
and full-rank, the DSF $(Q(s), P(s))$ can be uniquely factorized from the
transfer functions, which is assumed to be identifiable, and thereof from the
input-output data. To guarantee $P(s)$ to be diagonal, we need to impose
constrains on $(A, B)$ in the identification of state-space models. The
constrains rely on the following proposition in \cite{Hayden2016a}.

\begin{proposition}[P-Diagonal Form 1 \cite{Hayden2016a}]
  \label{prop:P-diag-form-Hayden}
  Any DSF $(Q,P)$ with $P$ square, diagonal and full rank has a realization with
  $A_{12}$, $A_{22}$, $B_1$ and $B_2$ from \eqref{eq:ss-sys} (no noise, and
  $C = \bm{I & 0}, D = 0$) partitioned as follows:
  \begin{equation}
    \label{eq:p-diag-AB}
    \left[
      \begin{array}{c@{\;\;}|c@{\,}c@{\;\;}}
        A_{12} && B_1 \\ \hline
        A_{22} && B_2
      \end{array}
    \right] =
    \left[
      \begin{array}{c@{\;\;\;}c@{\;\;}|c@{\,}c@{\;\;}c}
        \hat{c} & 0 && 0 & 0 \\
        0 & \times && 0 & B_{1_{22}}\\
        \hline
        \hat{a} & \times && \hat{b} & 0\\
        0 & \times && 0 & 0
      \end{array}
    \right]
  \end{equation}
  where $\times$ denotes an unspecified entry. The following is a canonical
  realization of $V = A_{12} (sI - A_{22})^{-1} B_2 + B_1$:
  \begin{equation}
    \label{eq:p-diag-V-realization}
    \begin{pmatrix}
      \hat{a}, & [\hat{b}\;\; 0], &
      \begin{bmatrix}
        \hat{c} \\ 0
      \end{bmatrix}, &
      \begin{bmatrix}
        0 & 0 \\ 0 & B_{1_{22}}
      \end{bmatrix}
    \end{pmatrix}
  \end{equation}
  where $\hat{a} \coloneqq \diag(\alpha_1, \dots, \alpha_{p_{11}})$, $\hat{b}
  \coloneqq \diag(\beta_1, \dots, \beta_{p_{11}})$ and $\hat{c} \coloneqq
  \diag(\gamma_1,\dots,\gamma_{p_{11}})$, $p_{22} = \dim(B_{1_{22}})$, $p_{11} = p -
  p_{22}$ and where $(\alpha_i, \beta_i, \gamma_i, 0)$ is a minimal realization of
  $V(i,i)$ in controllable canonical form.
\end{proposition}

To clarify how to integrate identifiability constraints, we need to review
the implementation of the SBL. Consider the noiseless case, where we
allow $\sigma^2 \rightarrow 0$. Using results from linear algebra, we have the
following expression for ${\mu}_\mathrm{w}$ and $\Sigma_\mathrm{w}$:
\begin{equation}
  \label{eq:SBL-posterior--mean-cov-2}
  \begin{array}{l@{\:}l}
    {\mu}_\mathrm{w} &= {\Gamma}^{1/2} \left( \Phi {\Gamma}^{1/2}
                           \right)^{\dag} \mathrm{y}\\
    \Sigma_\mathrm{w} &= \left[ I - {\Gamma}^{1/2} \left( \Phi
                        {\Gamma}^{1/2} \right)^{\dag} \Phi \right] {\Gamma},
  \end{array}
\end{equation}
where $(\cdot)^{\dag}$ denotes the Moore-Penrose pseudoinverse. It is clear to
see that the hyperparameters going to zero will lead to their corresponding
$\mathrm{w}$ elements being zero.  At the beginning of each EM step, we first
prune the data matrices $\Phi, \mathrm{y}$ by checking elements in ${\Gamma}$
that are smaller than the threshold of zero.  To deal with noisy cases, the
associated code by \cite{Wipf2007} performs the SVD decomposition on
$\Phi {\Gamma}$, denoted by $U S V^T$, and the estimation of
${\mu}_\mathrm{w}$ is updated by
${\Gamma} V \big(S (S^2 + \hat{\sigma}^2 I + \epsilon I)^{-1} \big) U^T
\mathrm{y}$, where $\hat{\sigma}^2$ is the estimated value in the previous step
and $\epsilon$ is a fixed value close to zero, e.g., $10^{-16}$. During the EM
iterations, the sizes of $\Phi$ and $\mathrm{y}$ will be significantly reduced
(depending on the sparsity of $\mathrm{w}$), which guarantees the computational
efficiency.

The objective of network identifiability integration is to guarantee the
resultant $P$ being square, diagonal and full-rank. Apparently, one necessary
condition is that $m = p$. However, this is not enough. Furthermore, the key
result used is the \emph{P-Diagonal Form} from \cite{Hayden2016a}. Let us start
with the simple case, where each input perturbs each output node independently,
i.e.  $B = [\diag(b_1,\dots,b_p) \;\; 0]^T$. Instead of putting $\kvec(B)$ in
$\mathrm{w}$, we only include $[b_1 \;\; \cdots \;\; b_p]$ and modify the data
matrix $\Phi$ correspondingly. The alternative is to keep the form of
$\mathrm{w}$ in \eqref{eq:SBL-likelihood}, while in the implementation of the EM
algorithm for SBL, we set the hyperparameters to zeros that correspond to zeros
in $B$. The general case is more complicated due to the unknown dimension
$p_{22} = \dim(B_{1_{22}})$ in Proposition~\ref{prop:P-diag-form-Hayden}. It is
integrated in the same way that setting the hyperparameters to zeros that
corresponds to zeros in \eqref{eq:p-diag-AB} (note that
$\hat{a},\hat{b},\hat{c}$ are diagonal with dimension $p - p_{22}$).
Unfortunately, we have not fully understood the importance of $p_{22}$. It is
possible that there exist multiple $p_{22}$'s leading to diagonal $P$'s, which
may or may not affect network reconstruction. It also has not yet been clear on
how to select $p_{22}$, which ``fortunately'' has at most $p$ choices ($p_{22}$
is an integer). However, this issue becomes serious if the size of network
(i.e. $p$) is huge.

Now we summarize everything and present the whole algorithm. The parameters for
initial states $m_0, R_0$ are treated as unknown deterministic variables, which
can be estimated in each M-step by $m_0^k = \hat{x}_{0|N}, R_0^k =
M_{0|N}$. This estimate comes from the study on optimizing
$\mathcal{Q}(\theta, \theta')$ given in
\eqref{eq:expect-compl-data-log-likelihood}, e.g., see \cite{Gibson2005}.
Parameters $\sigma^2, A, B$ and hyperparameters ${\gamma}$ are estimated via
another inner-loop EM procedure for the SBL. $A$ and $B$ are thought of as
random variables, whose estimations are given by
\eqref{eq:SBL-posterior--mean-cov}.  The whole method is given as
Algorithm~\ref{alg:SBL-EM-alg}.
\begin{algorithm}
  \caption{SBL embedded in the EM framework}
  \label{alg:SBL-EM-alg}
  \begin{algorithmic}[1]
    \State Initialize $\mathrm{w}^0, \gamma^0, (\sigma^2)^0$ and $m_0^0,
    R_0^0$. Choose the dimension of state space $n$.
    \While{1}
        \State Compute $\hat{x}_{k|N}, P_{x|N}, M_{x|N}$ and \eqref{eq:Expectation-xk-xk}-\eqref{eq:Expectation-xk-1-uk-1}
        ($k=0,\dots,N$) via Kalman smoothers in
        Lemma~\ref{lemma:expectation-items-for-compl-data-log-likeli}
        using the given parameters $\mathrm{w}^k, \gamma^k, (\sigma^2)^k, m_0^k, R_0^k$.
        \State Update $x_0^{k+1} = \hat{x}_{0|N}$, $R_0^{k+1} = M_{0|N}$.
        \State Compute $\Phi, \mathrm{y}$ using state estimations from Step~3.
        \State Perform another EM procedure \eqref{eq:SBL-alg-E-step},
        \eqref{eq:SBL-alg-M-step} to obtain $\mathrm{w}^{k+1}$ and
        $(\sigma^2)^{k+1}$, in which set the elements of $\gamma$ that
        correspond to zeros in \eqref{eq:p-diag-AB} to be zeros.
        \State \textbf{break} if parameter estimate converges.
    \EndWhile

    \State Compute $Q(s), P(s)$ by definition.
  \end{algorithmic}
\end{algorithm}

\section{Numerical Examples}
\label{sec:numerical-examples}

The empirical study is performed on random stable sparse state-space models
\eqref{eq:ss-sys} with $C = [I\;\; 0], D = 0$, from which the sparse DSF models
are derived. See \cite{Yue2017a} for the way to generate such sparse stable
state-space models.  Here we do not include such sparse networks whose
realizations' $A$ matrices are not sparse, which cannot be tackled by the
proposed method. The dimension of the networks is set to $p = 40$, i.e., the
dimension of output variables, and the dimension of states is set to $n =
100$.
The choice of sampling frequency is another issue deserving our particular
attention, due to \emph{system aliasing} (see \cite[chap.~3]{yue2018}).  The
sampling frequency for each system is chosen to be at least 10 times larger than
the critical frequency of system aliasing. This rule of thumb can mostly
guarantee that the discrete-time and continuous-time DSFs share the same network
structure.

To guarantee network identifiability from data, in the simulations, $p$-variate
Gaussian i.i.d. is used as input signals to drive each output node
separately. It implies the case in consideration is
$B = [I_{40 \times 40}\;\; 0]^T$. The general case is not included that inputs
may drive outputs via hidden states under the guarantee that $P$ is diagonal. It
is due to the technical difficulty of generating an appropriate $B$
automatically when $A$ is random. The ``general'' case might also be barely
interesting in practice: one cannot know how to design the inputs that perturb
nodes via hidden states without knowing the hidden state variables or
models. However, it should be included in future work to complete the test.

The proposed method runs on 50 random networks, where the assumed dimension of
states is set to 110 (the ground truth is 100). In the performance benchmark, we
focus on Boolean structures of $Q$ ($P$ has been known to be diagonal and $H$ is
not quite interesting in applications). The reconstruction results are
summarized in Table~\ref{tbl:results-dsf-disc-cont}, where three columns of SNR
show the averaged \emph{Precision} (i.e. the percentage of correct links in
results) and the averaged \emph{TPR} (i.e. the percentage of links of the ground
truth shown in results). The regularization parameter for ``TSM-NR'' is selected
approximately by balancing between the values of {Precision} and {TPR}. The
value of \emph{failure} denotes the percentage of results that fail to show
sparse structures, which is computed as the averaged percentage of results for
three cases of different SNRs with Precision < 5\%. We exclude these ``failed''
networks when we compute the averaged values of Precision and TPR. As shown in
Table~\ref{tbl:results-dsf-disc-cont}, the state-space based method (``SSM-NR'')
provides a better way to perform reconstruction of discrete-time DSFs than
identifying specific parametric models (``TSM-NR'').  The poor performance of
``TSM-NR'' for low SNRs is due to the restrictive choice of ARX (due to
difficulty on numerical optimization). The modeling uncertainty of the ARX
parametric method becomes significant and cannot be dealt with by assuming it as
noises when the SNR is low.  The existence of ``failed'' cases is probably due
to the random construction of random networks, which cannot be covered by the
proposed method but are difficult to be removed automatically in random model
generation.

\begin{table}[htb]
  \centering
  \caption{Reconstruction results of the proposed method (labeled as ``SSM-NR'')
    and the parametric method \cite{Yue2017a} (labeled as ``TSM-NR'') (rounded
    to zero decimals).}
  \begin{tabular}{c|c|r|r|r|r}
    \hline\hline
    & & \multicolumn{3}{c|}{SNR} &  \\ \cline{3-5}
    & & 0 dB & 20 dB & 40 dB & \raisebox{1.5ex}[0pt]{Failure} \\ \hline
    & Precision & 76\% & 83\% & 94\% & \\ \cline{2-5}
    \raisebox{1.5ex}[0pt]{SSM-NR} & TPR & 81\% & 78\% & 89\%
      & \raisebox{1.5ex}[0pt]{<\,4\%} \\ \hline
    & Precision & 40\% & 74\% & 81\% & \\ \cline{2-5}
    \raisebox{1.5ex}[0pt]{TSM-NR} & TPR & 60\% & 65\% & 83\%
      & \raisebox{1.5ex}[0pt]{0\%} \\ \hline
  \end{tabular}
  \label{tbl:results-dsf-disc-cont}
\end{table}

The parametric methods require to choose model orders of each element in
$Q,P,H$, which implies that a huge amount of parameters need to be determined in
model selection. This might be hardly practical using AIC, BIC, or
cross-validation, if we have little prior knowledge.  In the proposed method,
the DSFs are derived from state space models, which allow more complicated time
series models besides ARX (e.g., it allows the elements of $Q$ in the same row
share different poles). That is the main reason why the proposed method has
better performance. Moreover, in terms of model selection, the proposed method
demands only 1 parameter to be determined, i.e. the size of states. This value
could be easily determined by information criteria, cross-validation, or by
performing other state-space identification methods, e.g., the subspace method.

\section{Conclusions and Outlook}
\label{sec:conclusions}

This paper proposes an algorithm to reconstruct sparse dynamic networks using
the EM algorithm embedded with sparse Bayesian learning.  The algorithm
reconstructs the DSFs by identifying their feasible state-space
realizations. Kalman smoothers are used to provide state estimation.  To
guarantee network identifiability, which ensures unique reconstruction of
network structures, samples are measured from experiments that perturb each
output independently. The identifiability conditions are integrated in the
inner-loop EM iteration for sparse Bayesian learning.

This work shows outlook for future studies. We consider special cases of
state-space models, in the perspective of system identification. The extension
of the work to general LTI state-space models relies on further studies of
network identifiability on the generalized DSF models. A line of research that
is, albeit interesting and import, might be challenging.  Another issue is to
ensure sparse DSFs. The proposed method use a heuristic approach, which assumes
that the sparse DSFs under consideration have sparse realizations. However, the
remaining class of DSFs that is excluded in this work, has not been quantified
and shown to have measure zero within the parameter space of LTI dynamical
systems. A better way to reconstruct networks could be to impose sparsity
directly on the DSFs models.

\appendix[Dynamical Structure Functions]

The procedure to define the DSF \eqref{eq:dsf-sys} from \eqref{eq:ss-sys} mainly
refers to \cite{Goncalves2008,Chetty2015}. Without loss of generality, suppose
that $C$ is full row rank (see \cite{Chetty2015} for a general $C$). Create the
$n\!\times\!n$ state transformation $T = \bm{C^T & E}^T$, where
$E \in \mathbb{R}^{n\times (n-p)}$ is any basis of the null space of $C$ with
$T^{-1} = \bm{\bar{E} & E}$ and $\bar{E} = C^T(C C^T)^{-1}$. Now we change the
basis such that $z = Tx$, yielding $\hat{A} = TAT^{-1}$, $\hat{B} = TB$,
$\hat{C} = CT^{-1}$, $\hat{D} = D$, $K = TK$, and partitioned commensurate with
the block partitioning of $T$ and $T^{-1}$ to give
\begin{equation*}
  \begin{array}{r@{\:}c@{\:}l}
    \bm{z_1(t_{k+1}) \\ z_2(t_{k+1})} &=&
    \begin{bmatrix}
      \hat{A}_{11} & \hat{A}_{12} \\ \hat{A}_{21} & \hat{A}_{22}
    \end{bmatrix}
    \bm{z_1(t_k) \\ z_2(t_k)} +
    \bm{\hat{B}_1 \\ \hat{B}_2} u(t_k) \\
                  && + \bm{\hat{K}_1 \\ \hat{K}_2} e(t_k), \\
    y(t_k) &=& \bm{I & 0} \bm{z_1(t_k) \\ z_2(t_k)} + D u(t_k) + e(t_k).
  \end{array}
\end{equation*}
Introduce the shift operator $q$ and solve for $z_2$,
yielding $qz_1(t_k) = W(q) z_1(t_k) + V(q) u(t_k) + L(q) e(t_k) $, where
$W(q) = \hat{A}_{11} + \hat{A}_{12}(qI - \hat{A}_{22})^{-1} \hat{A}_{21}$,
$V(q) = \hat{B}_{1} + \hat{A}_{12}(qI - \hat{A}_{22})^{-1} \hat{B}_{2}$, and
$L(q) = \hat{H}_{1} + \hat{A}_{12}(qI - \hat{A}_{22})^{-1} \hat{H}_{2}$.  Let
$D_{W}(q) = \diag(W(q))$ be a diagonal matrix function composed of the diagonal
entries of $W(q)$. Define $\hat{Q}(q) = (qI - D_W)^{-1}(W-D_W)$,
$\hat{P}(q) = (qI - D_W)^{-1}V$, and $\hat{H}(q) = (qI - D_W)^{-1}L$, yielding
\begin{math}
  z_1(t_k) = \hat{Q}(q) z_1(t_k) + \hat{P}(q) u(t_k) + \hat{H}(q) e(t_k).
\end{math}
Noting that $z_1(t_k) = y(t_k) - Du(t_k) - e(t_k)$, the DSF of \eqref{eq:ss-sys}
with respect to $y$ is then given by
\begin{equation}
  \label{eq:qph-general-dsf}
  \begin{array}{l@{\:}l}
    Q(q) &= \hat{Q}(q), \\
    P(q) &= \hat{P}(q) + (I-\hat{Q}(q))D,\\
    H(q) &= \hat{H}(q) + (I - \hat{Q}(q)).
  \end{array}
\end{equation}
Noting that the elements of $\hat{Q}, \hat{P}, \hat{H}$ (except zeros in the
diagonal of $\hat{Q}$) are all strictly proper, it is easy to see that $Q$ is
strictly proper and $P,H$ are proper. It has been proven in \cite{Chetty2015}
that the DSF defined by this procedure is invariant to the class of block
diagonal transformations used above, which implies it is a feasible extension of
the definition of DSFs given in \cite{Goncalves2008} for the particular class of
state-space models with $C = \bm{I & 0}, D = 0$.

\def\url#1{}      
\bibliographystyle{IEEEtran}
\bibliography{./ref/library}

\end{document}

%% file: supports/userdef-mathsym.tex
\newcommand{\mnorm}[1]{{\left\vert\kern-0.25ex\left\vert\kern-0.25ex\left\vert #1
    \right\vert\kern-0.25ex\right\vert\kern-0.25ex\right\vert}}    
\newcommand{\bm}[1]{\begin{bmatrix} #1 \end{bmatrix}}

\newcommand{\diag}{\operatorname{{diag}}}
\newcommand{\blkdiag}{\operatorname{{blkdiag}}}

\newcommand{\kvec}{\operatorname{{vec}}}

\newcommand{\Tr}{\operatorname{{Tr}}}

\newcommand{\argmax}[1]{\underset{#1}{\operatorname{arg\,\!max}\;}}
\newcommand{\argmini}[1]{\operatorname{arg}\,\!\underset{#1}{\operatorname{min}}\;}

